\documentclass[
reprint,
superscriptaddress,
 amsmath,amssymb,
 aps,
pra,
]{revtex4-2}

\usepackage{graphicx}
\usepackage{dcolumn}
\usepackage{bm}
\usepackage[utf8]{inputenc}
\usepackage{subcaption}
\usepackage{mhchem}
\usepackage{multirow}
\usepackage{placeins} 
\usepackage{makecell}
\usepackage{amsmath}
\usepackage{hyperref}
\usepackage{color,soul}
\usepackage{chemformula}
\usepackage{comment}
\usepackage{textcmds}
\usepackage{caption}
\usepackage{chemscheme}
\usepackage{amsfonts}
\usepackage{csquotes}
\usepackage{adjustbox}
\usepackage{booktabs}
\usepackage{float}

\begin{document}

\preprint{APS/123-QED}

\title{Materials Database from All-electron Hybrid Functional DFT Calculations}

\author{Akhil S. Nair$^*$}
\affiliation{ The NOMAD Laboratory at the Fritz Haber Institute of the Max Planck Society, Faradayweg 4-6, D-14195 Berlin, Germany}
\affiliation{Institute of Chemistry and Biochemistry, Freie University Berlin, Arnimallee 22, 14195 Berlin, Germany}
\author{Lucas Foppa}
\affiliation{ The NOMAD Laboratory at the Fritz Haber Institute of the Max Planck Society, Faradayweg 4-6, D-14195 Berlin, Germany}
\author{Matthias Scheffler}
\affiliation{ The NOMAD Laboratory at the Fritz Haber Institute of the Max Planck Society, Faradayweg 4-6, D-14195 Berlin, Germany}

\date{\today}

\begin{abstract}
Materials databases built from calculations based on density functional approximations play an important role in the discovery of materials with improved properties. Most databases thus constructed rely on the generalized gradient approximation (GGA) for electron exchange and correlation. This limits the reliability of these databases, as well as the artificial intelligence (AI) models trained on them, for certain classes of materials and properties which are not well described by GGA. In this paper, we describe a database of 7,024 inorganic materials presenting diverse structures and compositions generated using hybrid functional calculations enabled by their efficient implementation in the all-electron code FHI-aims. The database is used to evaluate the thermodynamic and electrochemical stability of oxides relevant to catalysis and energy related applications. We illustrate how the database can be used to train AI models for material properties using the sure-independence screening and sparsifying operator (SISSO) approach.
\end{abstract}

\maketitle

\section{Background and Summary}
Materials serve as the cornerstone of critical economic sectors, including transportation, health, information technology and energy. Hence, the discovery of materials with improved properties is crucial. Materials databases (e.g., Materials Project \cite{jain2013commentary}, OQMD \cite{kirklin2015open}, AFLOW \cite{curtarolo2012aflow}) constructed with high-throughput calculations based on density functional approximations (DFAs), have played a key role in accelerating materials discovery. These databases enable the construction of artificial intelligence (AI) models for material properties  \cite{batatia2023foundation, merchant2023scaling}. However, the usefulness of such databases, and consequently the applicability of AI models trained on them, are constrained by the accuracy of the underlying theoretical assumptions used to generate the data. As a result, these models might fail when deployed to explore materials and properties that are not well described by the chosen DFA. For example, electron exchange correlation (XC) functionals under the generalized gradient approximation (GGA) such as the Perdew-Burke-Ernzerhof (PBE) \cite{perdew1996generalized}, face accuracy limitations for estimating the electronic properties, especially for systems with localized electronic states such as transition-metal oxides \cite{zhang2018performance,wang2006oxidation}. Recently, high-throughput studies using the meta-GGA functional SCAN \cite{schmidt2022dataset} or its regularized variants \cite{kingsbury2022performance} and non-local, range-separated hybrid functional HSE06 \cite{kim2020band,liu2024high} have been used for constructing materials databases which, to some extent, address the accuracy limitations of GGA. However, many of these studies focused on specific material properties (e.g., band gaps) or system types (e.g., binary solids). Additionally, most of them employed plane-wave basis sets with pseudopotential approximation for core electrons, which may face accuracy and transferability challenges across materials with diverse crystal structures \cite{lejaeghere2016reproducibility, mazdziarz2024uncertainty}. All-electron calculations using beyond-GGA DFAs can provide more reliable data, and databases constructed from such calculations could play a crucial role in developing AI models with enhanced generalizability.

Here, we present a database of 7,024 materials constructed from  all-electron DFT calculations with hybrid functional for XC with a focus on oxides relevant for catalysis and energy related applications. The construction of this database has been enabled by the scalability-improved implementation in FHI-aims which allows to perform hybrid functional calculations at feasible costs for a large number of materials \cite{sebastian}. We also illustrate how the database facilitates the development of AI models for material properties using the symbolic-regression based Sure-Independence Screening and Sparsifying Operator (SISSO) \cite{ouyang2018sisso} approach, which identifies the key parameters correlated with material properties, making the model interpretable.

\section{Methods}

\begin{figure*}[ht]
   \centering
   \includegraphics[width=\textwidth]{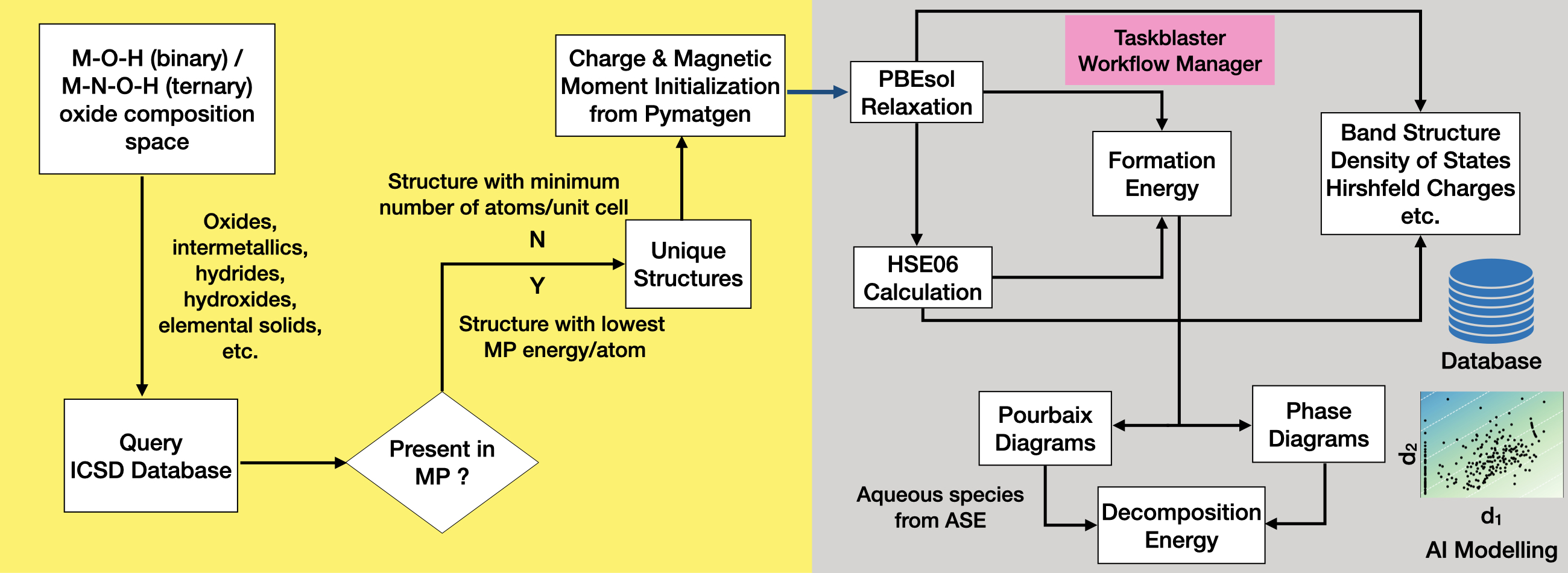}
   \caption{Computational workflow employed for curating the database. $d_1$ and $d_2$ correspond to the descriptors derived from SISSO model.
 }
   \label{Fig: workflow}
\end{figure*}
The database consists of 7,024 materials. We considered $M$-O-H ($M$-$N$-O-H) as the reference chemical system for selecting binary (ternary) materials as we are interested in evaluating their thermodynamic and electrochemical stability. This involves oxides ($M$-O/$M$-$N$-O), intermetallics ($M$-$N$), hydrides ($M$-H/$N$-H/$M$-$N$-H), hydroxides ($M$-O-H/$N$-O-H/$M$-$N$-O-H) and elemental solids ($M$/$N$). The initial crystal structures for these materials are queried from the Inorganic Crystral Structure Database (ICSD, v2020). Since ICSD consists of a large number of materials with the same composition (duplicate entries or polymorphs), filtering of entries for which the ICSD-id is associated with at least one Materials Project ID (MP-id, v2023.11.1) is done based on the lowest energy/atom criteria according to MP (GGA/GGA+U) data. In case there is no MP-id found for a formula, the ICSD entry with the lowest number of atoms in the unit cell is considered. Notably, we imposed no restrictions on unit cell sizes or crystal prototypes, resulting in structures with up to 616 atoms per unit cell. A schematic representation of the computational workflow is provided in Fig. \ref{Fig: workflow}.
\\
\\
For the selected structures,  geometry optimizations are performed with the PBEsol functional \cite{perdew2008restoring} as it provides an accurate estimation of lattice constants of solids \cite{csonka2009assessing}. Using the structures optimized with the PBEsol, we performed HSE06 energy evaluation and electronic structure calculation as HSE06 gives more accurate electronic properties  \cite{chevrier2010hybrid, garza2016predicting}. Our previous studies confirmed that HSE06 provides only slight improvements in lattice constants with respect to GGA functionals \cite{zhang2018performance,nair2024materials}. Hence, we used HSE06 single-point calculations on structures optimized with the PBEsol XC.   For both PBEsol and HSE06, outputs including the electronic band structure, density of states, and Hirshfeld charges are also computed. All the calculations are performed using the all-electron code FHI-aims using numerically atom-centered orbtial (NAO) basis sets with "light" settings. A computational workflow based on the Taskblaster framework \cite{taskblaster} is used to automate the multiple tasks involved in creating he data. A convergence criterion of $\mathrm{10^{-3}}$ $\mathrm{eV/\AA}$ is considered for forces. For all the potentially magnetic structures (ie., either labelled as magnetic in MP or containing elements such as Fe, Ni, Co, etc.), spin-polarized calculations are performed. Among the initially queried materials, HSE06 electronic structure calculations did not converge for 198 materials and hence are excluded from the database.
\vspace{-5pt}
\section{Data Records}
All electronic structure data from FHI-aims calculations can be accessed from the NOMAD archive  \url{https://nomad-lab.eu/prod/v1/gui/user/datasets/dataset/id/sijndGsxT32FufFQfqFvvw}. The SQLite3 ASE databases can be downloaded from the figshare repository \url{10.6084/m9.figshare.28375538}. In Table 1, we provide a description of the processed data. All the properties computed using PBEsol and HSE06 are made available in tabular format.
\\

\begin{table}[!ht]
\centering
\caption{\textbf{Data fields in tabulated form.}}
\begin{tabular}{lp{0.7\textwidth}}
\hline
\textbf{Field} & \textbf{Description} \\
\hline
Formula & Chemical composition of the material \\
Reduced formula & Chemical composition normalized by formula units \\
$N_{\text{atoms}}$ & Number of atoms in the unit cell of the material \\
$E_{\text{PBEsol}}$ & PBEsol calculated total energy  \\
$E_{\text{HSE06}}$ & HSE06 calculated total energy \\
$E_{gap,\text{PBEsol}}$ & PBEsol calculated band gap \\
$E_{gap,\text{HSE06}}$ & HSE06 calculated band gap \\
$\Delta H_{f,\text{PBEsol}}$ & PBEsol calculated formation energy \\
$\Delta H_{f,\text{HSE06}}$ & HSE06 calculated formation energy \\
$\Delta H_{d,\text{PBEsol}}$ & Decomposition energy from PBEsol convex hull phase diagram \\
$\Delta H_{d,\text{HSE06}}$ & Decomposition energy from HSE06 convex hull phase diagram \\
$\Delta G^{\text{OER}}_{pbx^,\text{PBEsol}}$ & Decomposition energy from PBEsol Pourbaix diagram at pH=0 and applied potential $U=1.23$ V \\
$\Delta G^{\text{OER}}_{pbx^,\text{HSE06}}$  & Decomposition energy from HSE06 Pourbaix diagram at pH=0 and applied potential $U=1.23$ V \\
\hline
\end{tabular}
\end{table}

\begin{figure*}[ht!]
\captionsetup{justification=justified,singlelinecheck=false}
  \begin{subfigure}{0.75\textwidth}
    \includegraphics[width=\linewidth]{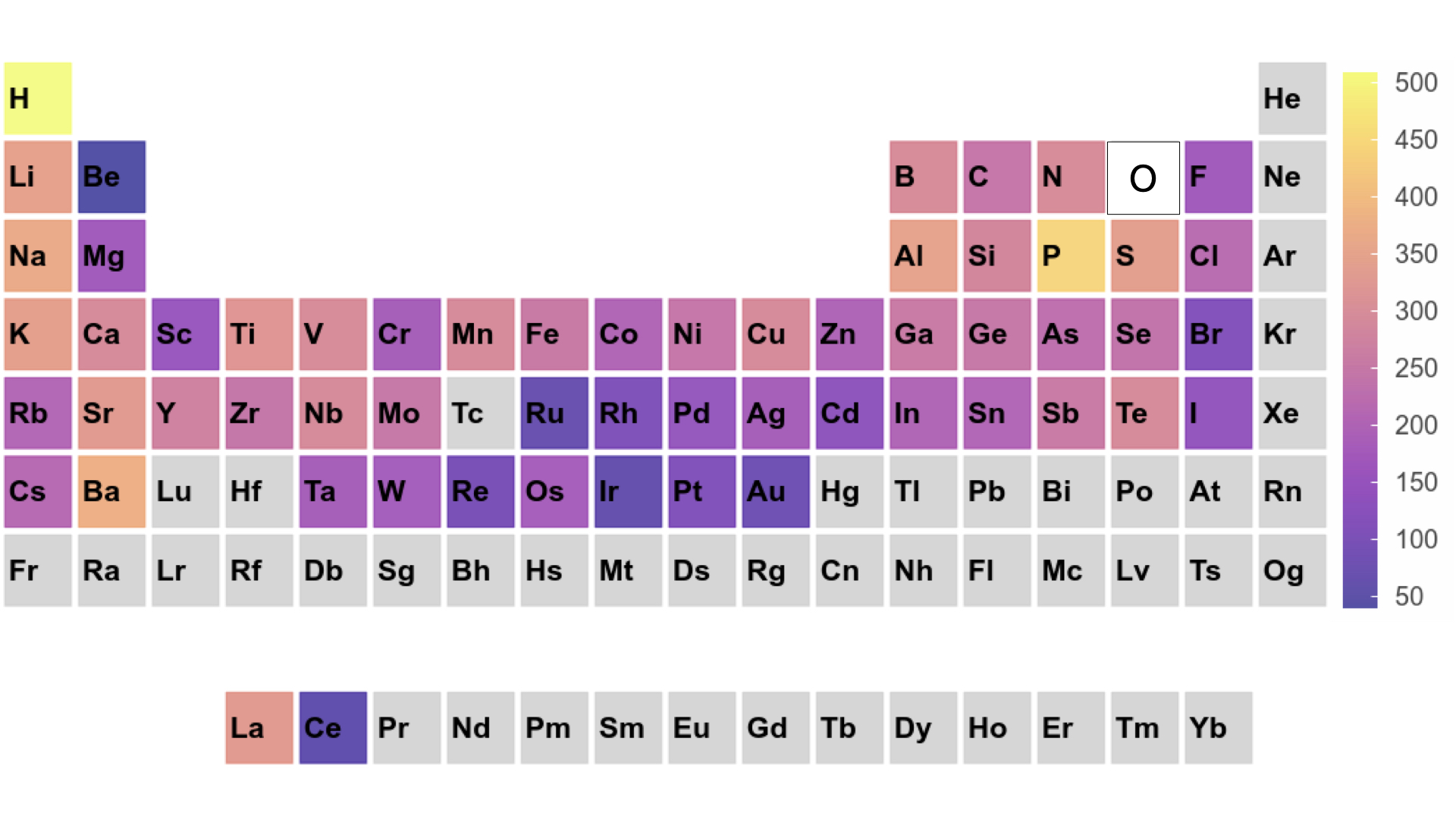}
    \caption{} 
  \end{subfigure}  
\captionsetup{justification=justified,singlelinecheck=false}
  \centering
  \begin{subfigure}{0.3\textwidth}
    \includegraphics[width=\linewidth]{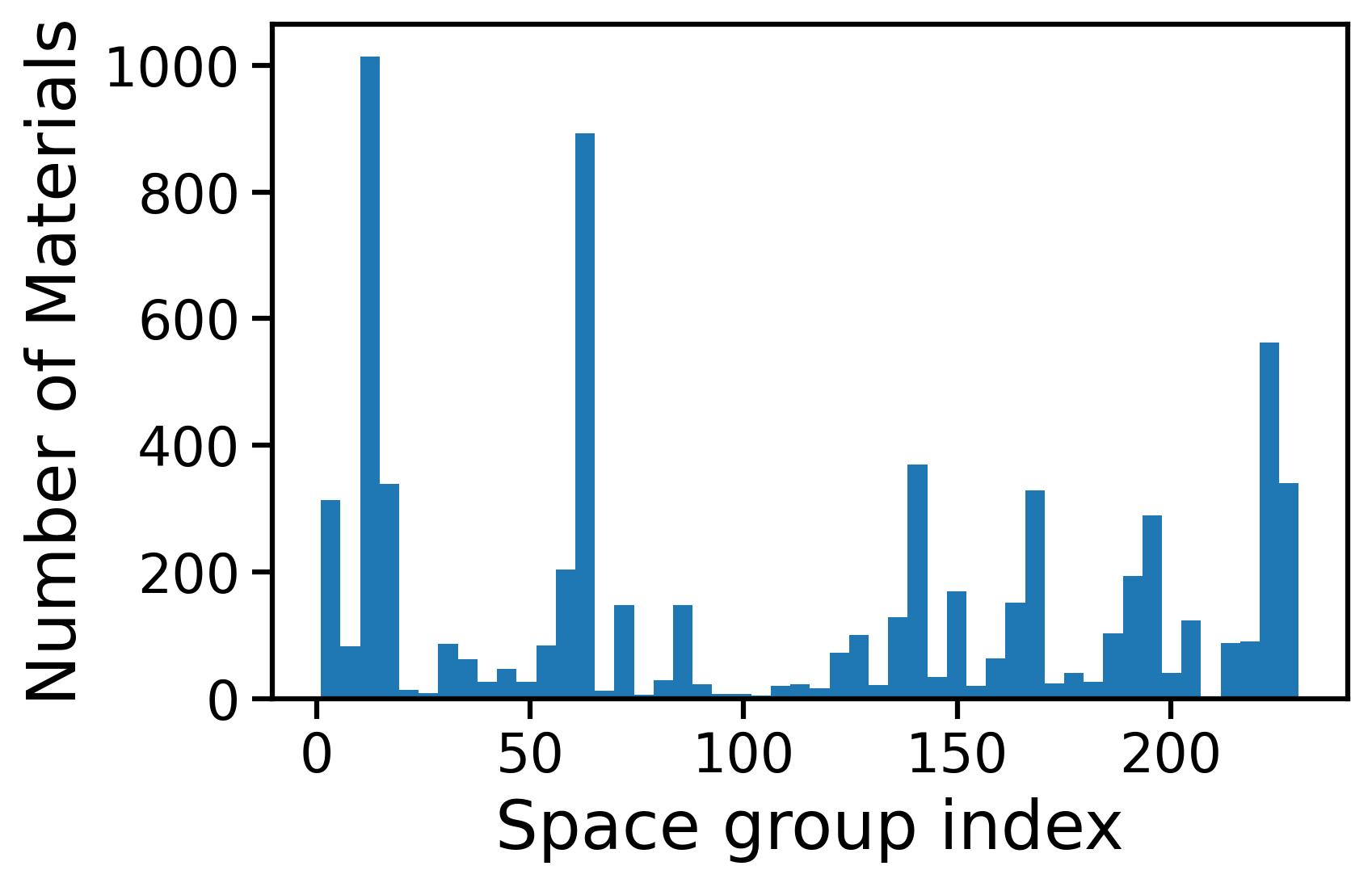}
    \caption{} 
  \end{subfigure}  
  \begin{subfigure}{0.35\textwidth}
    \includegraphics[width=\linewidth]{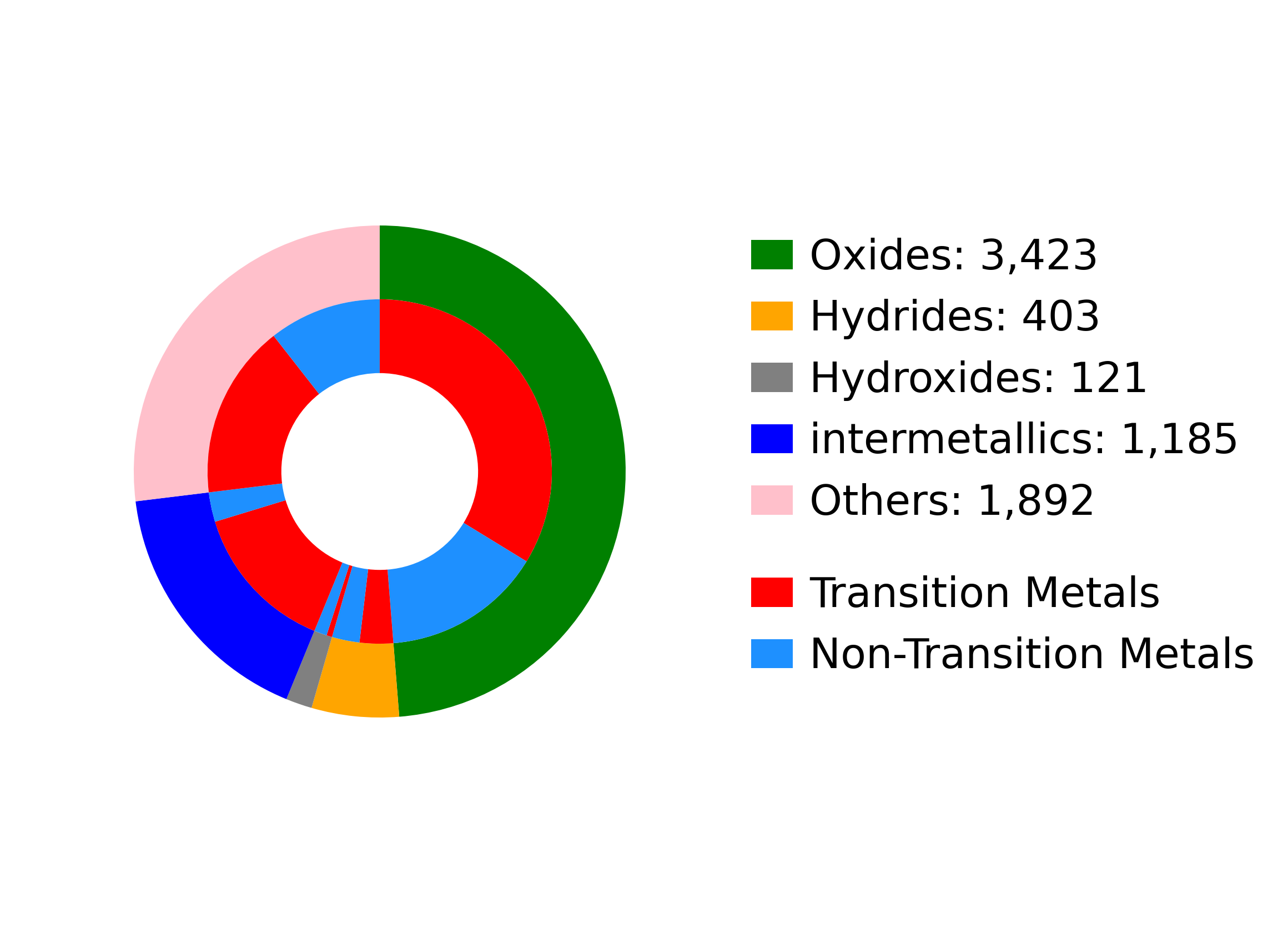}
    \caption{} 
  \end{subfigure}
  \begin{subfigure}{0.3\textwidth}
    \includegraphics[width=\linewidth]{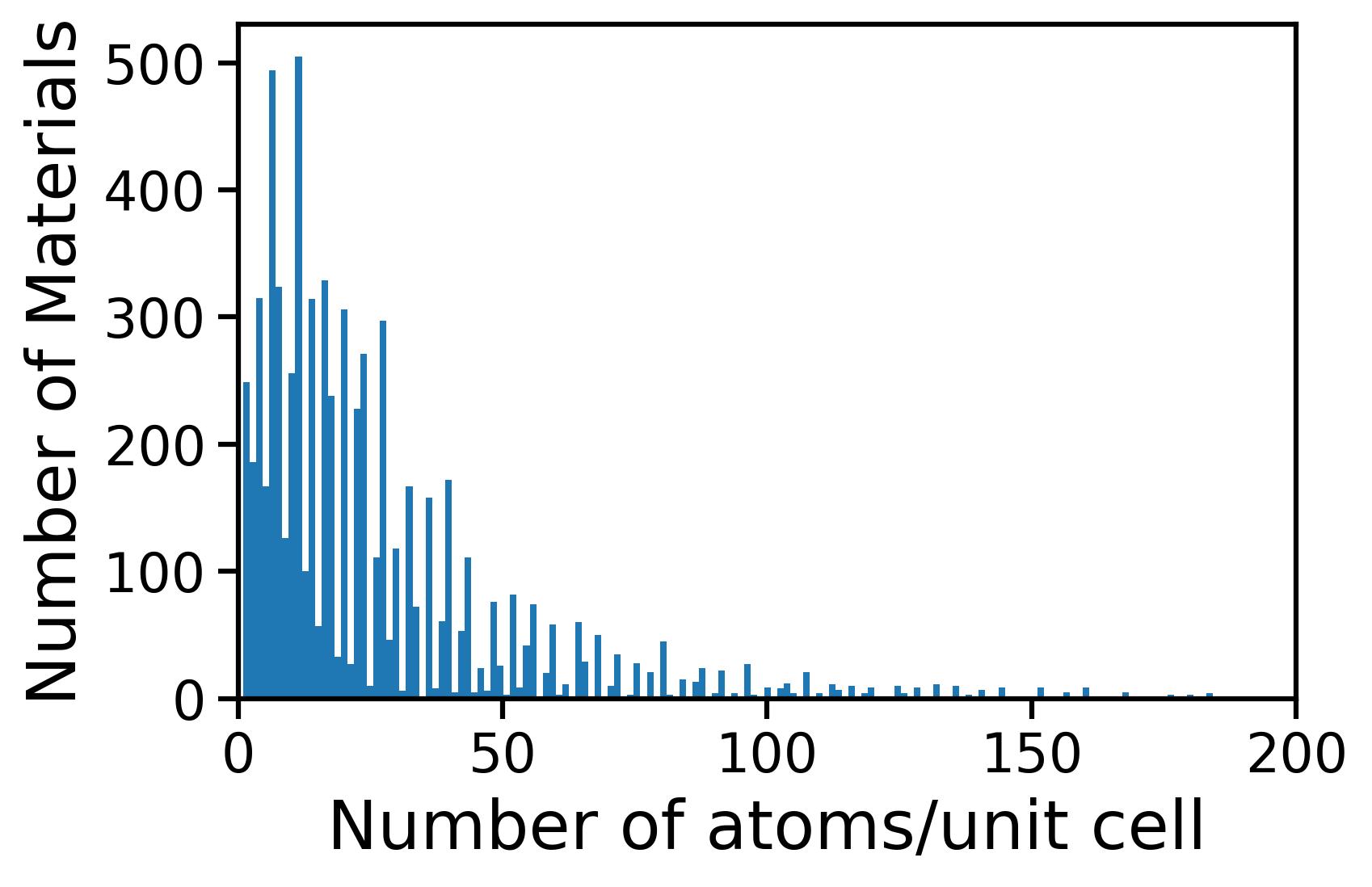}
    \caption{} 
  \end{subfigure}
  \caption{(a) Periodic table style heatmap showing the frequency of elements constituting the materials present in the database, distribution of (b) space group indexes, (c) material categories and (d) number of atoms in the unit cell across the materials in the database.} 
  \label{Fig:details_database}
\end{figure*}

Fig. \ref{Fig:details_database}a reflects the chemical diversity of the materials in the database. The materials of the database present elements covering the majority of the periodic table except actinides, noble gases and elements with the atomic number $\geq$ 80. Fig. \ref{Fig:details_database}b shows the distribution of space groups (determined using spglib \cite{togo2024spglib} with tolerance of $10^{-5}$ \AA) across the database. The most frequent spacegroups observed are $P2_1/c$, $Pnma$, $Fm$-$3m$, $C2/m$, and $P6_3mc$. As shown in  \ref{Fig:details_database}c, we observe that a significant proportion of materials contain at least one transition metal. Fig. \ref{Fig:details_database}d illustrates the distribution of the number of atoms per unit cell (limited to 200 for ease of visualization). While the majority of materials have 25 or fewer atoms in their unit cell, 14\% of materials have unit cells with more than 50 atoms. 

\section{Technical Validation}

\begin{figure*}[!htbp]
  \begin{subfigure}{0.49\textwidth}
    \includegraphics[width=\linewidth]{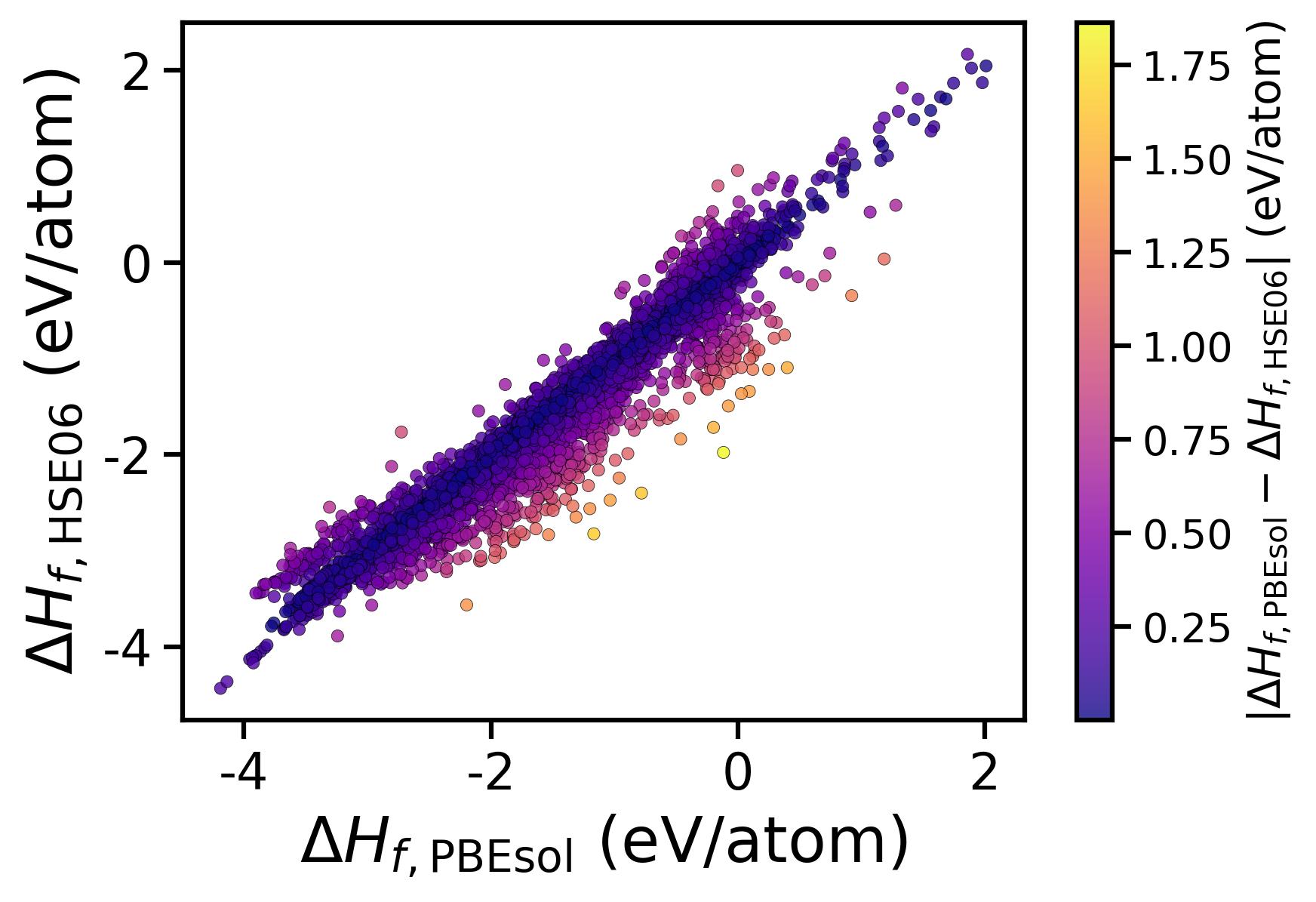}
    \caption*{(a)} \label{Fig:3a}
  \end{subfigure}%
  \begin{subfigure}{0.48\textwidth}
    \includegraphics[width=\linewidth]{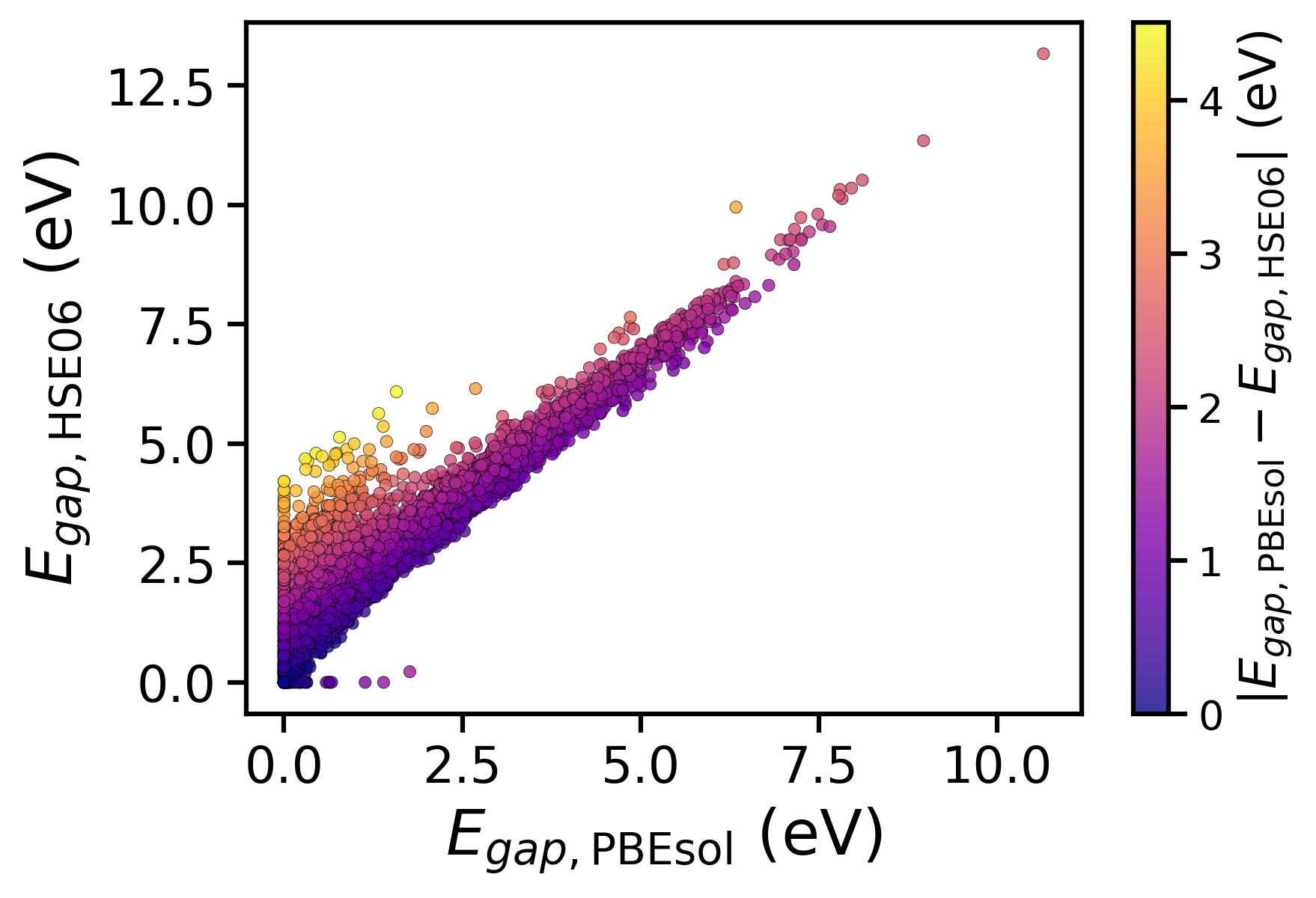}
    \caption*{(b)} \label{Fig:3b}
  \end{subfigure}%
  \caption{Comparison of (a) formation energies and (b) band gaps between PBEsol and HSE06 for the materials in the database. The color gradient represents the absolute difference in the property values between the methods.}
  \label{Fig:form_gap_compare}
\end{figure*}

We first compared the formation energies and band gaps computed using PBEsol and HSE06. The formation energies are calculated by considering bulk phases for the elements except oxygen for which gaseous \ce{O2} is considered as the reference phase. In general, HSE06 provides lower formation energies compared to PBEsol as shown in  Fig. \ref{Fig:form_gap_compare}a. A mean absolute deviation (MAD) of 0.15 eV/atom is found between the formation energies calculated by the two methods. Compared to formation energies, a higher disparity (MAD = 0.77 eV) is observed in the band gaps estimated by PBEsol and HSE06, as evident from Fig. \ref{Fig:form_gap_compare}b with HSE06 showing a shift in band gap values toward higher ranges. This is expected owing to the well-known underestimation of band gaps by GGA methods which is partially corrected by HSE06. For 342 materials, PBEsol estimates a metallic character whereas HSE06 provides a band gap value $>=$0.5 eV. We also performed a benchmarking against experimental band gap data for binary systems curated by Pedro \textit{et al.} \cite{borlido2019large}. For the 121 materials common to both our dataset and theirs (determined by MP-id), PBEsol yields a mean absolute error (MAE) of 1.35 eV, which improves by over 50\% with HSE06 (0.62 eV). Note that a comprehensive benchmarking of formation energies and band gaps against experimental measurements is precluded by the limited availability and significant uncertainties of existing experimental data.  This limitation is further compounded by the challenges high-throughput simulations face in accurately capturing magnetic ordering and defect-related effects under experimental conditions \cite{horton2019high,nieminen2009issues}.

\begin{figure*}[!htbp]
  \begin{subfigure}{0.45\textwidth}
    \includegraphics[width=\linewidth]{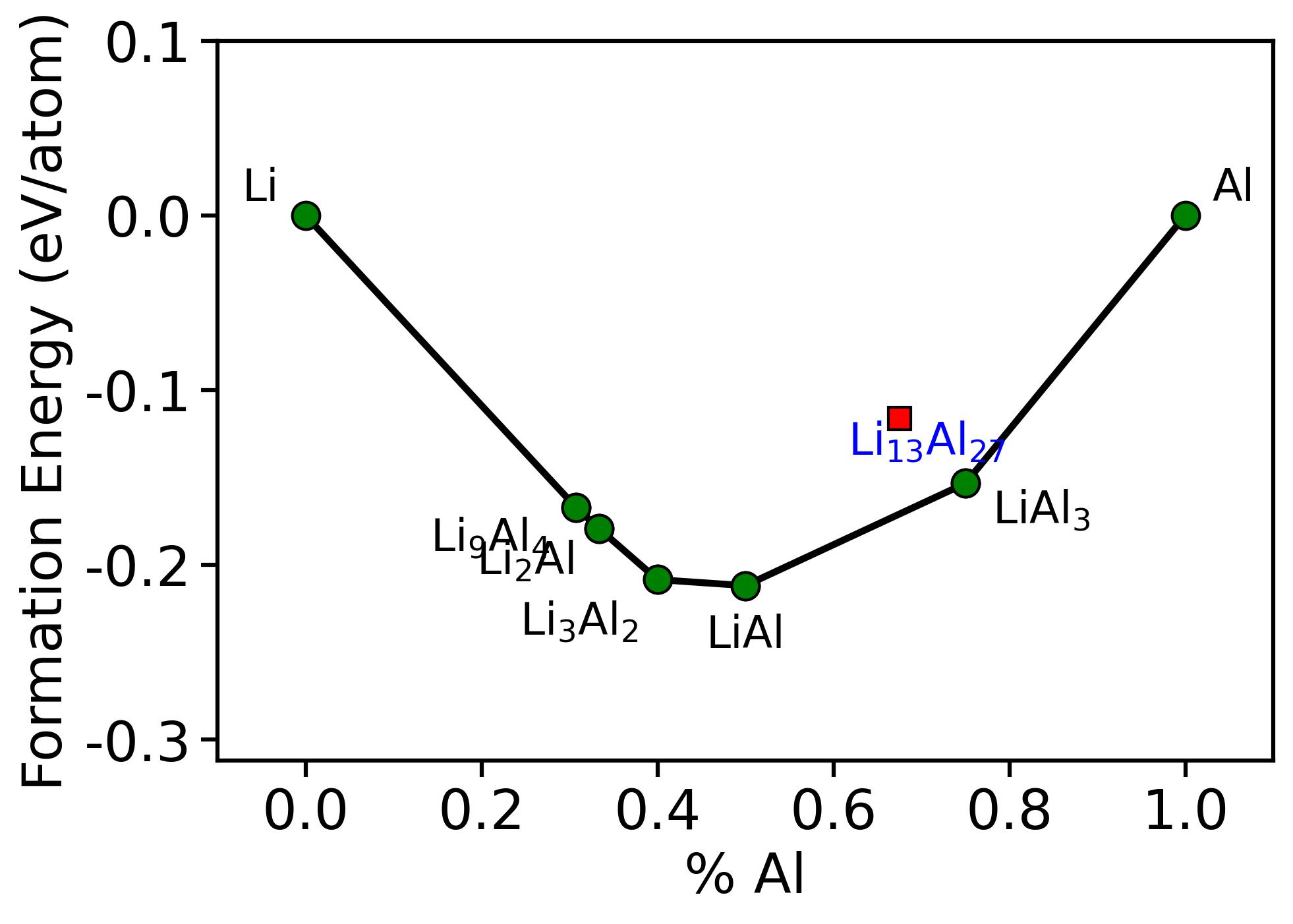}
    \caption*{} \label{Fig:10a}
  \end{subfigure}%
  \begin{subfigure}{0.45\textwidth}
    \includegraphics[width=\linewidth]{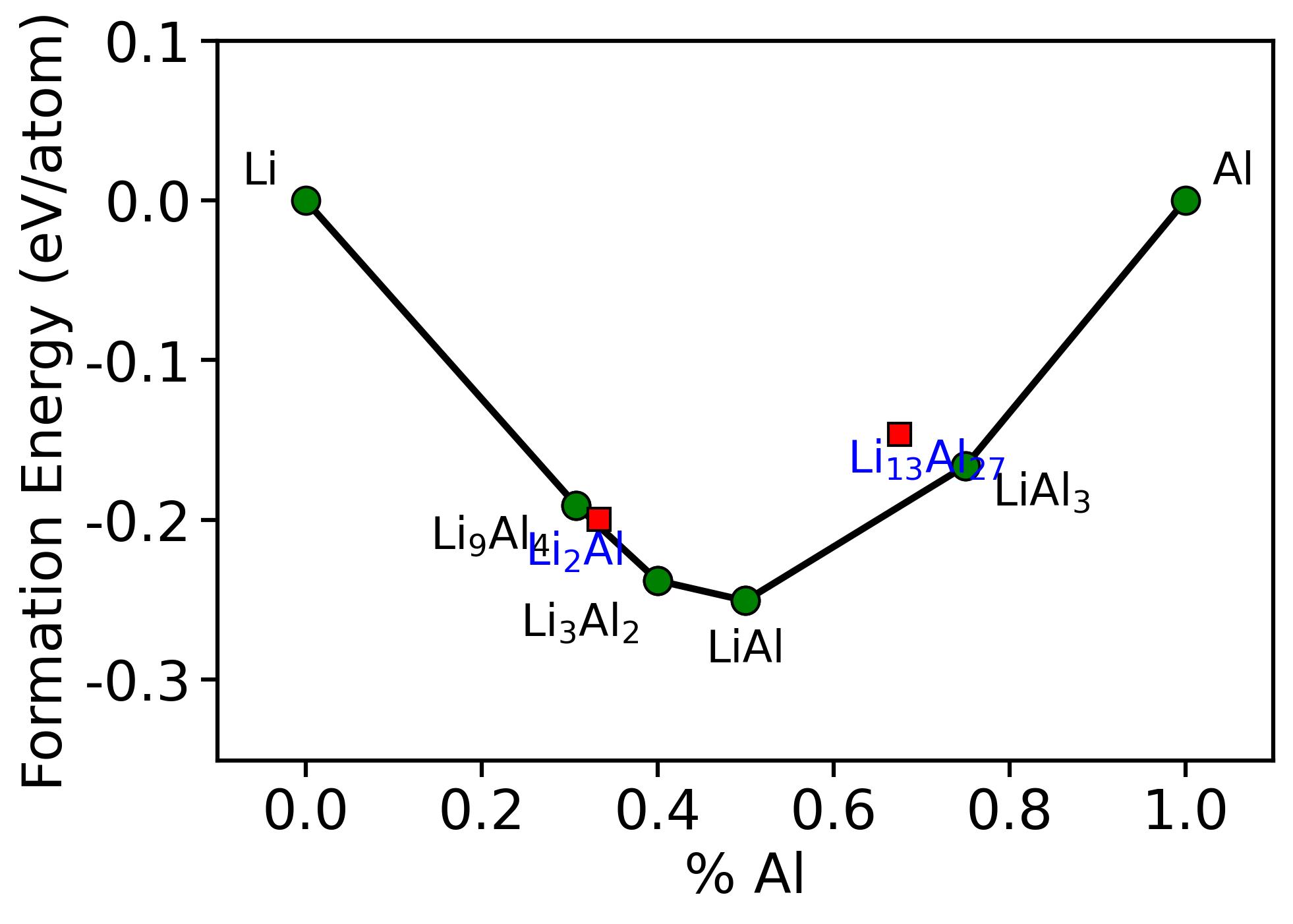}
    \caption*{} \label{Fig:10b}
  \end{subfigure}%
  \\
  \begin{subfigure}{0.45\textwidth}
    \includegraphics[width=\linewidth]{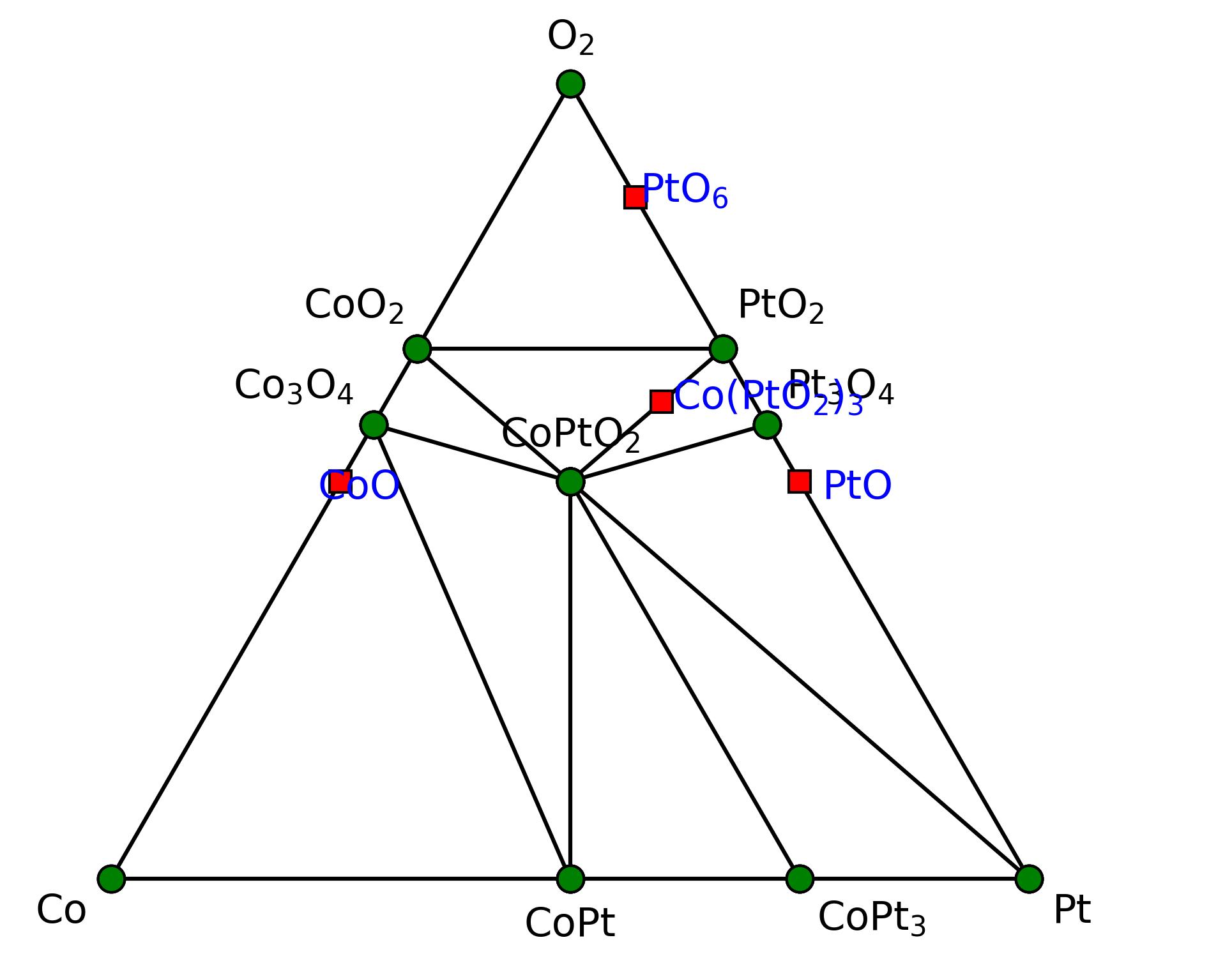}
    \caption*{} \label{Fig:10c}
  \end{subfigure}
    \begin{subfigure}{0.45\textwidth}
    \includegraphics[width=\linewidth]{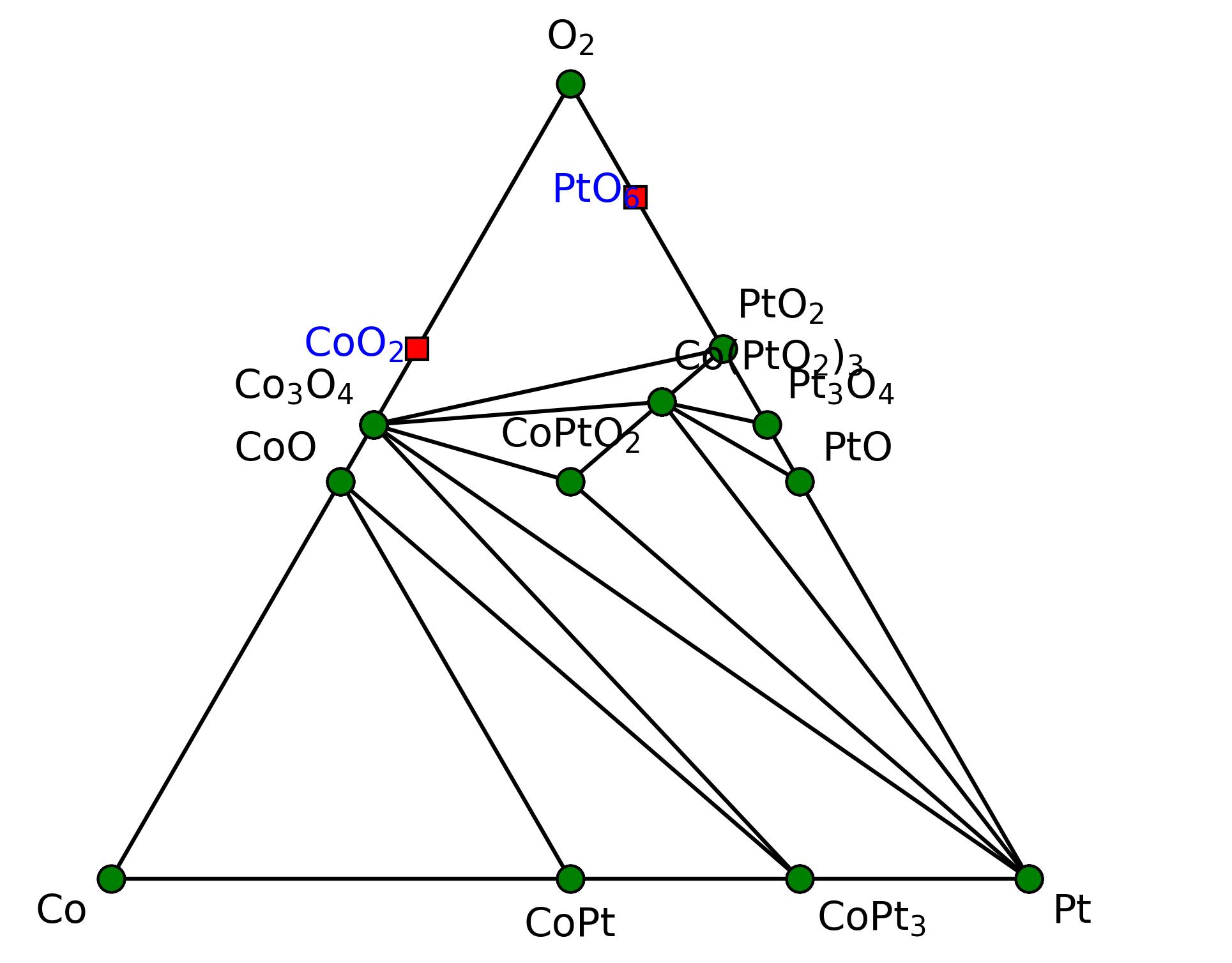}
    \caption*{} \label{Fig:10d}
  \end{subfigure}
  \caption{Convex hull phase diagram of Li-Al (upper panel) and Co-Pt-O (lower panel) chemical systems calculated with PBE (left) and HSE06 (right). The stable and unstable phases (energy above hull $>$0) are denoted by green circles and red squares, respectively.}
  \label{Fig:CPD}
\end{figure*} 

\begin{figure*}[!htbp]
  \begin{subfigure}{0.49\textwidth}
    \includegraphics[width=\linewidth]{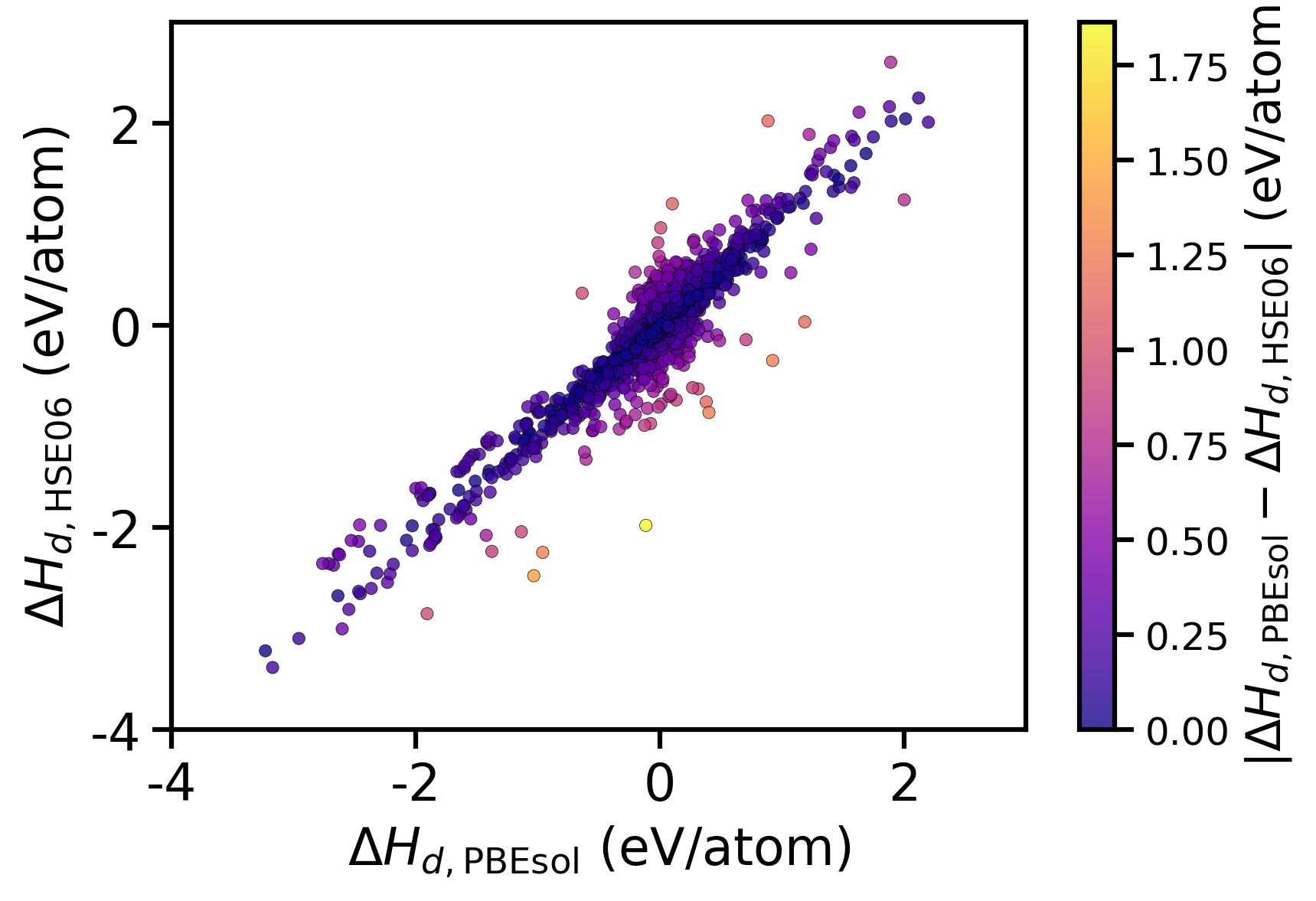}
    \caption*{(a)} \label{Fig:5a}
  \end{subfigure}%
  \begin{subfigure}{0.48\textwidth}
    \includegraphics[width=\linewidth]{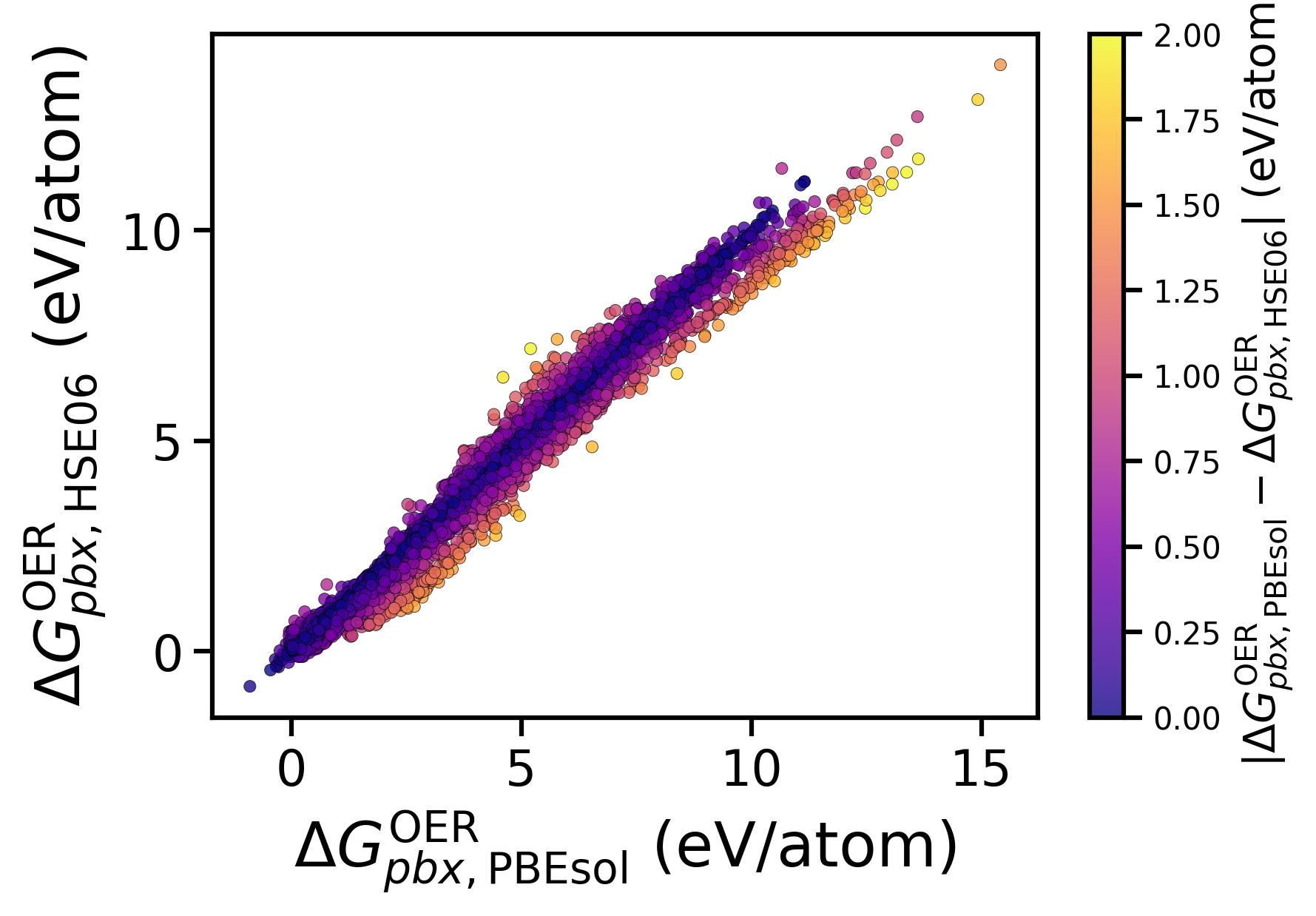}
    \caption*{(b)} \label{Fig:10b}
  \end{subfigure}%
  \caption{Comparison of PBEsol vs. HSE06 decomposition energies from convex hull phase diagrams (left) and Pourbaix diagrams (right) for the materials in the database.}
  \label{Fig:decomp_compare}
\end{figure*}

To understand whether the accuracy differences in formation energies translate to thermodynamic stability, we constructed the convex hull phase diagrams (CPDs) of the materials in the database. We choose Li-Al and Co-Pt-O as representative chemical systems for binary and ternary compositions and their phase diagrams are shown in Fig. \ref{Fig:CPD}. Distinct CPDs are obtained by PBEsol and HSE06 for these systems. For example, \ce{Li2Al} is stable with PBEsol but slightly unstable by 4 meV with HSE06 and similarly, \ce{Co(PtO3)2} is unstable with PBEsol by 11 meV but stable with HSE06. The critical decomposition reactions (decomposition in the CPD with the highest positive reaction energy) are determined for all the materials present in the database along with the associated decomposition energy ($\Delta H_d$) as a quantitative metric of phase stability. For Li-Al, all the decomposition reactions are identical for both PBEsol and HSE06. However, different critical decomposition reactions are identified for \ce{Pt3O4}, \ce{Co3O4} and \ce{CoPtO2} from the Co-Pt-O CPD. Some phases such as \ce{PtO6} are found to have the same critical decomposition reaction but with significantly differing  $\Delta H_d$ values for PBEsol (0.17 eV/atom) and HSE06 (0.5 eV/atom).

\begin{figure*}[!htbp]
  \begin{subfigure}{0.48\textwidth}
    \includegraphics[width=\linewidth]{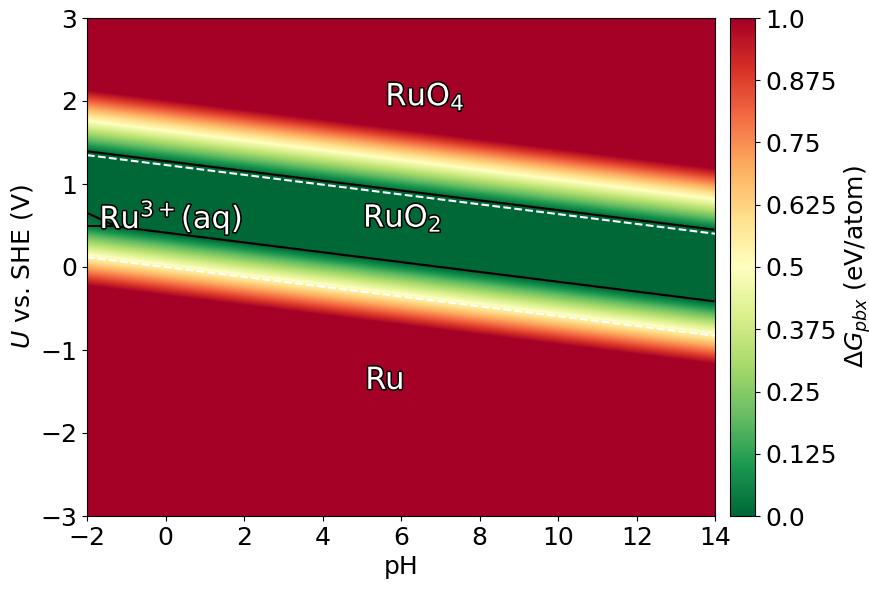}
    \caption*{} \label{Fig:5a}
  \end{subfigure}%
  \begin{subfigure}{0.48\textwidth}
    \includegraphics[width=\linewidth]{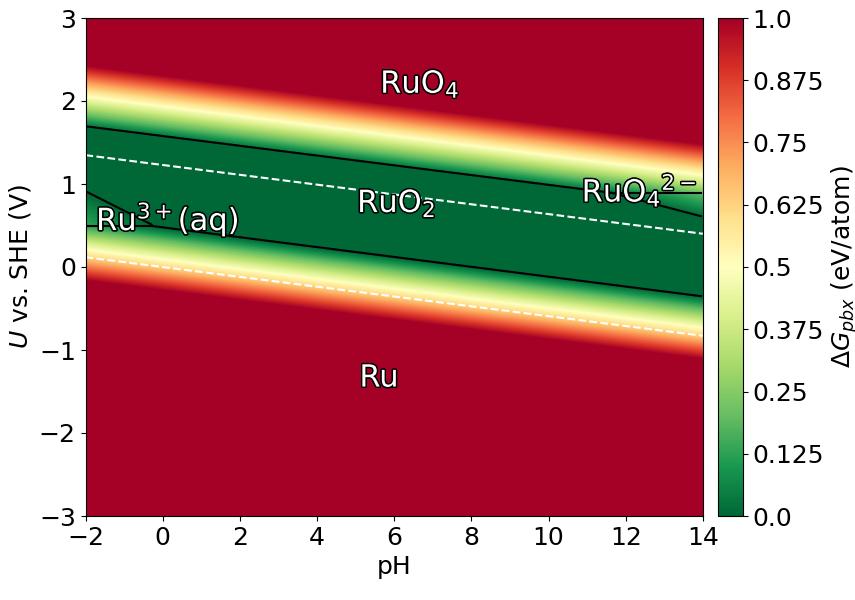}
    \caption*{} \label{Fig:10b}
  \end{subfigure}%
  \\
  \begin{subfigure}{0.48\textwidth}
    \includegraphics[width=\linewidth]{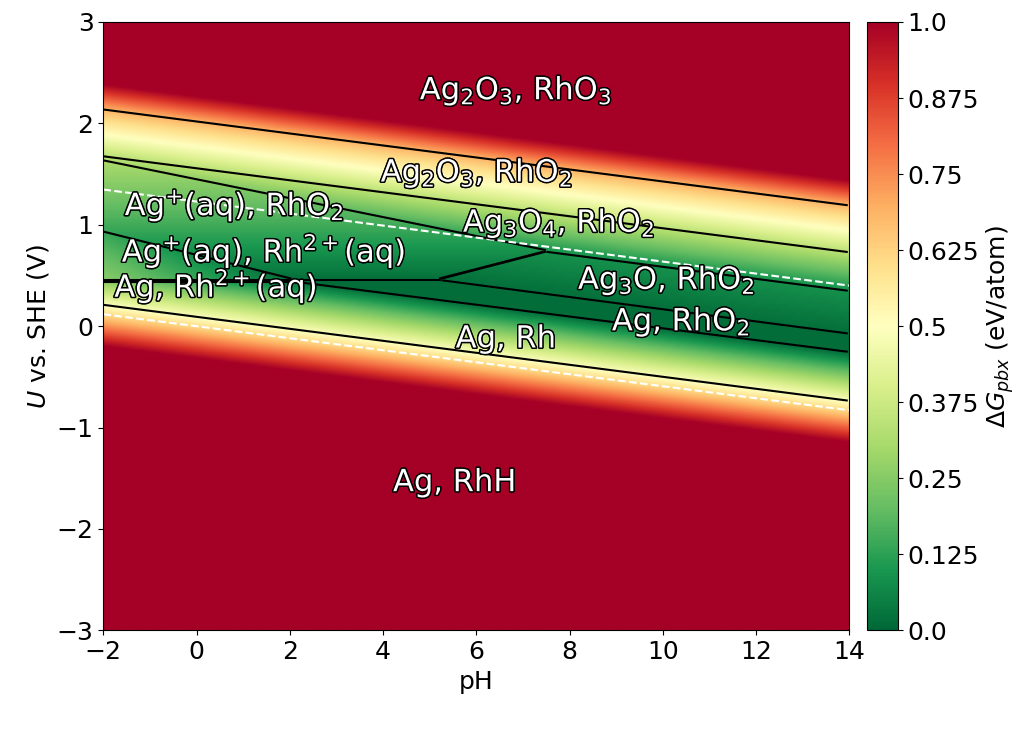}
    \caption*{} \label{Fig:10c}
  \end{subfigure}
    \begin{subfigure}{0.485\textwidth}
    \includegraphics[width=\linewidth]{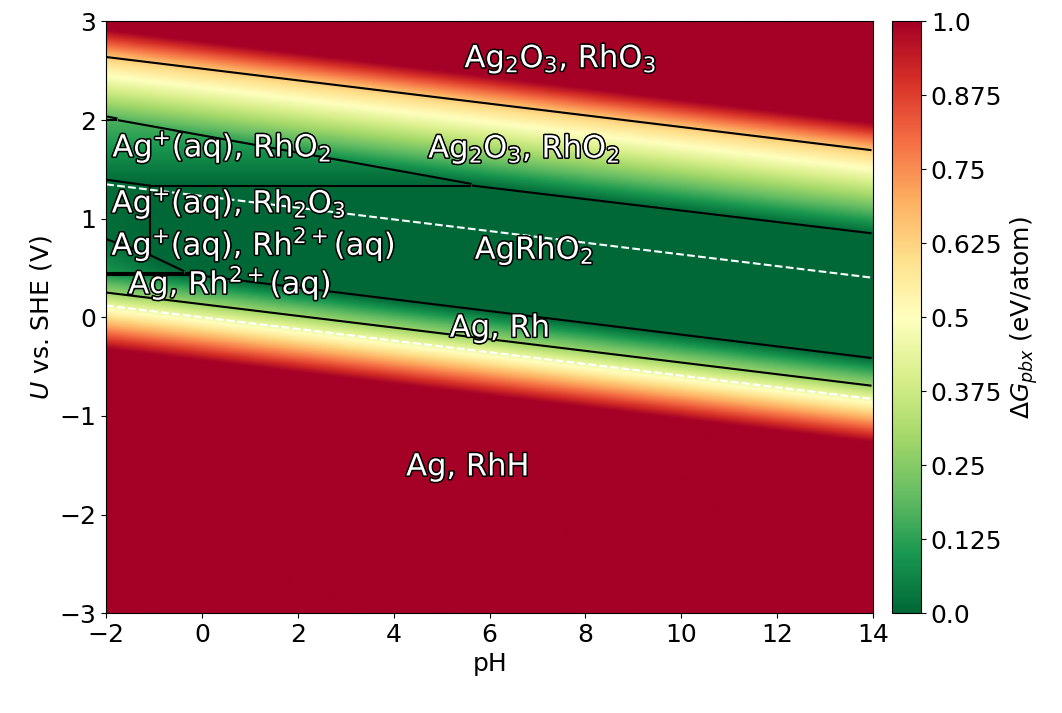}
    \caption*{} \label{Fig:10d}
  \end{subfigure}
  \caption{Ru (upper panel) and Ag-Rh (lower panel) Pourbaix diagram calculated using PBEsol (left) and HSE06 (right).  An aqueous ion concentration of $10^{-6}$ M and temperature of 298.15 K are considered for the construction of Pourbaix diagrams. The chemical potentials of ions are retrieved from ASE database. The green and red regions indicate stable and unstable domains of the pH-$U$ coordinate space with respect to the reference solid. The stability domain of \ce{H2O} is shown with white dashed lines.}
  \label{Fig:pourbaix}
\end{figure*}

To analyze the phase stability prediction trends between PBEsol and HSE06, we plot the $\Delta H_d$ distributions obtained from both the methods in Fig. \ref{Fig:decomp_compare}a. 78\% and 75\% of the considered materials are found having $\Delta H_d \leq 50$ meV/atom for PBEsol and HSE06 respectively. The higher $\Delta H_d$ values for the remaining materials despite them being synthesized (ie., associated with an ICSD-id) could be attributed to synthetic conditions which are not accounted for by the zero-Kelvin DFT stability prediction. The general tendency of HSE06 estimating lower formation energies does not lead to more materials being stable in the corresponding CPD, as many of these materials decompose into binary or ternary compounds.  It is important to note that this analysis is limited by the absence of experimental data for certain phases and also by the exclusion of polymorphs, both of which are crucial for constructing more accurate CPDs.

\begin{figure}[h!]
   \centering
   \includegraphics[width=\textwidth]{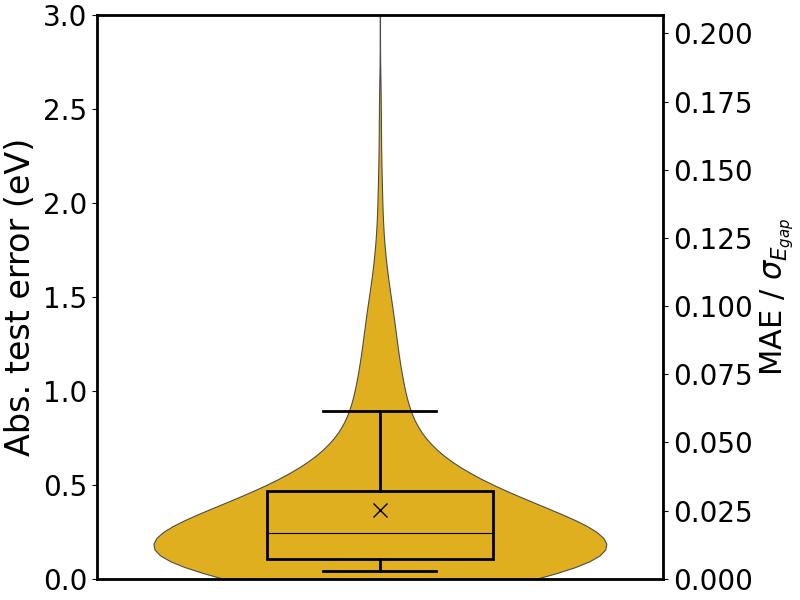}
   \caption{Distriution of nested cross validation test errors (primary y-axis) of the SISSO model for the $E_{gap,\text{HSE06}}$ prediction.  In the secondary y-axis, the ratio of mean absolute error and standard deviation of $E_{gap,\text{HSE06}}$  is shown. The area between the upper and lower edges of the black box represents the interquartile range of errors and the whiskers from bottom to top correspond to 10 percentile, median and 90 percentile errors. The mean and median of absolute errors are indicated by the cross mark and horizontal line inside the interquartile range, respectively.
 }
   \label{Fig:violin}
\end{figure}

In addition to phase stability, we have also determined the electrochemical stability of the materials as this is relevant for their potential applicability to areas such as electrocatalysis, corrosion resistance, batteries etc. This is carried out by constructing Pourbaix diagrams \cite{pourbaix1966atlas,persson2012prediction} which provide information regarding the material's stability across different redox conditions within a potential-pH coordinate space, electrochemical decomposition pathways, passivation and corrosion regions. Representative PBEsol and HSE06 calculated Pourbaix diagrams for Ru and \ce{Ag-Rh} systems by considering \ce{RuO2} and \ce{AgRhO2} as the target materials of interest are given in Fig. \ref{Fig:pourbaix}. These oxides have been identified to be stable as part of our previous study employing the DFT-HSE06 method \cite{nair2024materials}. For both systems, we observe qualitative differences in the Pourbaix diagrams obtained across the methods. In the case of Ru, the stability domain of both \ce{RuO2} and \ce{Ru3+} are enlarged in the HSE06 Pourbaix diagram compared to PBEsol. An additional phase \ce{RuO4^{2-}} is also present in the HSE06 diagram. In this study, we limit our focus to analyzing the acid-stability (hence used equivalent to electrochemical stability from hereafter) of material under conditions (pH=0 and applied potential $U$=1.23 V) relevant to oxygen evolution reaction (OER), an important reaction involved in electrocatalytic water splitting.  As a quantitative metric of a material's acid-stability, we computed the Pourbaix decomposition free energy  ($\Delta G_{pbx}^{\mathrm{OER}}$) at these conditions. \ce{RuO2} has a $\Delta G_{pbx}^{\mathrm{OER}}$ of -0.06 eV/atom and -0.26 eV/atom with PBEsol and HSE06, respectively, indicating both the methods predict \ce{RuO2} as acid-stable though to different extents. The Ag-Rh Pourbaix diagrams generated using PBEsol and HSE06 exhibit differences in both the stable phases present and the pH-$U$ stability range of the common phases between the two methods. A notable difference is that PBEsol predicts \ce{AgRhO2} to be acid-unstable with $\Delta G_{pbx}^{\mathrm{OER}}$ = 0.21 eV/atom, whereas HSE06 predicts it to be stable with $\Delta G_{pbx}^{\mathrm{OER}}$ = -0.01 eV/atom. A comparison of $\Delta G_{pbx}^{\mathrm{OER}}$ calculated using PBEsol and HSE06 is provided in Fig. \ref{Fig:decomp_compare}b for all the materials for which Pourbaix diagrams are constructed. A MAD of 0.27 eV/atom is observed between the two methods. This indicates that the disparity in predicting the electrochemical stability is much higher compared to the phase stability which is expected due to the increased number of redox reactions under aqueous conditions. Under a stability criterion of $\Delta G_{pbx}^{\mathrm{OER}} \leq $ 0.1 eV/atom, 222 materials are acid-stable with PBEsol which is 255 for HSE06. 

We also developed AI models based on the database. Specifically, we focused on predicting the HSE06 band gaps of materials using SISSO, an interpretable AI method that identifies descriptors correlated with material properties. For this, a set of primary features, ie., basic material parameters which could be potentially correlated with the target property of interest are collected. These include elemental properties (composition-weighted), as well as structural and electronic properties determined using PBEsol. A dataset of 3091 materials with HSE06 band gaps $\geq$ 0.05 eV is considered for the model training. For accessing the model performance, we performed a nested cross validation (NCV) where the model validation and selection are done in the inner loop and model performance estimation is done on the outer loop. The SISSO model thus obtained is:


\begin{equation}
\begin{split}
E^{\text{SISSO}}_{g,\text{HSE06}} &= c_0 + 
a_0 \left[\frac{\langle N_{VAC} \rangle - \langle N_{VAL} \rangle}{\langle N_{VAL} \rangle \langle R_{VAL} \rangle} \right] + \\ 
&\quad a_1 \left[ \frac{\sqrt{\langle AN \rangle}}{\langle EN \rangle \langle R_{s} \rangle} \right] + \\ 
&\quad a_2 \left[\frac{(\langle EA \rangle)}{\langle AN \rangle}- E_{gap,\text{PBEsol}} \right] \\ 
\end{split}
\end{equation}

where $E_{g,\mathrm{PBEsol}}$ is PBEsol calculated band gap. $N_{VAL}$ is the number of valence orbitals, $R_{s}$ is the s-orbital radii, $R_{VAL}$ is the valence orbital radii, $EA$ is the electron affinity, $EN$ is the electronegativity, $AN$ is the atomic number, all computed as composition based weight averaged (indicated by $\langle \rangle$)  properties. The coefficients of the SISSO model are $\mathrm{c_0=2.45}$ eV, $\mathrm{a_0=-0.07 \AA eV}$, $\mathrm{a_1=-0.08 \AA eV^2}$ and $\mathrm{a_2= -1.05}$. 
Fig. \ref{Fig:violin} illustrates the distribution of test errors from NCV scheme, showing that the model achieves reasonable accuracy, with 90\% of errors falling within the 7.5\% of the ratio between mean test errors and the standard deviation of the band gaps in the dataset. This indicates that AI models trained on this database can achieve reliable HSE06 band gap predictions. 

The database presented in this work includes properties related to stability and electronic structure, evaluated at the HSE06 level of theory while assuming that PBEsol provides reasonable geometries for the material classes considered. However, for studying certain properties—such as defects or achieving more accurate band edge alignments—consideration of HSE06-relaxed geometries may be necessary. Additionally, our formation energy calculations do not account for temperature and pressure effects, which could be important for a more realistic phase diagrams under experimental conditions. The database is expected to serve as a reference for further investigations, including these refinements. Future improvements in the quality of materials data could also be enabled by employing higher-accuracy electronic structure methods beyond hybrid functionals, such as GW approaches.
\vspace{8pt}
\section{Code Availability}
All processed data and supporting scripts can be found in figshare \url{10.6084/m9.figshare.28375538} and the Gitlab repository \url{https://gitlab.com/akhilsnair/mat_database}. 

\bibliography{apssamp}
\end{document}